\documentclass[twocolumn,showpacs,preprintnumbers,amsmath,amssymb]{revtex4}
\usepackage{amssymb}
\usepackage{graphicx}
\usepackage{amsmath}
\usepackage[dvips]{epsfig}

\makeatletter
\def\btt#1{\texttt{\@backslashchar#1}}
\DeclareRobustCommand\bblash{\btt{\@backslashchar}}
\makeatother

\begin{document}

\preprint{Manuscript}
\title{Manipulating thermal conductivity through substrate coupling}
\author{Zhixin Guo, Dier Zhang and Xin-Gao Gong$^{\ast }$}
\affiliation{MOE Key Laboratory for Computational Physics, and Surface Physics Laboratory
(National Key), Fudan University, Shanghai 200433, China}
\email{xggong@fudan.edu.cn}
\date{\today }

\begin{abstract}
We report a new approach to the thermal conductivity manipulation --
substrate coupling. Generally, the phonon scattering with substrates
can decrease the thermal conductivity, as observed in recent
experiments. However, we find that at certain regions, the coupling
to substrates can increase the thermal conductivity due to a
reduction of anharmonic phonon scattering induced by shift of the
phonon band to the low wave vector. In this way, the thermal
conductivity can be efficiently manipulated via coupling to
different substrates, without changing or destroying the material
structures. This idea is demonstrated by calculating the thermal
conductivity of modified double-walled carbon nanotubes and also by
the ice nanotubes coupled within carbon nanotubes.\\\

\textbf{Keywords:} Thermal conductivity, manipulate, substrate
coupling, molecular dynamics
\end{abstract}

\maketitle

With the shrinkage of electronic devices to the nanoscale\cite%
{1,2,3,4} and the revival of thermoelectrics\cite{5,6,7}, thermal
transport property of nanomaterials has become more and more
significant. So far, great efforts have been done on
finding/synthesizing new nanomaterials with particularly high/low
thermal conductivities.\cite{7,8,9,10} Recently, some efforts have
taken headway on the subject of manipulating the thermal
conductivity via doping, adsorbing, or generating
defects.\cite{11,12,13,14,15} These processes can only reduce but
hardly enhance the thermal conductivity, since the phonon scattering
in a crystal lattice usually be less than that with the doping or
defects. Furthermore, most of these treatments largely destroy the
structure and thus the corresponding properties of nanomaterials,
which make them inapplicable in the nanodevices.

Very recently, the effects of coupling with substrates on thermal
conductivity were reported\cite{16,17,18}. In general, when a
conductive material is coupled with substrate, its thermal
conductivity is expected to be decreased owning to the additional
phonon scattering with the substrate as observed in the recent
experiments.\cite{16,17} In this letter, we find that, at certain
regions, the effect of phonon scattering can be suppressed and the
thermal conductivity of nanomaterials can be surprisingly increased
due to the coupling induced shift of phonon band to the low wave
vector. Based on this finding, we propose a new approach to thermal
conductivity manipulation--coupling with different substrates. This
approach naturally extends the capability of conventional treatments
on thermal conductivity without destroying the  structures of
materials. Since in the production of nanodevices, the conductive
nanomaterials are always placed on certain substrates, our approach
has great potential for advancing the performance of nanoelectronic
and thermoelectric devices.

In order to demonstrate this approach, we start with a coupled
Fermi-Pasta-Ulam(FPU) chain\cite{19,20} model. Then we use two
examples, thermal conductivity of both modified double-walled carbon
nanotubes (DWNTs) and coupled ice nanotubes (Ice-NTs)\cite{21}, to
demonstrate the applicability in real systems.

\begin{figure}[tbp]
\includegraphics[scale=0.3,angle=-90]{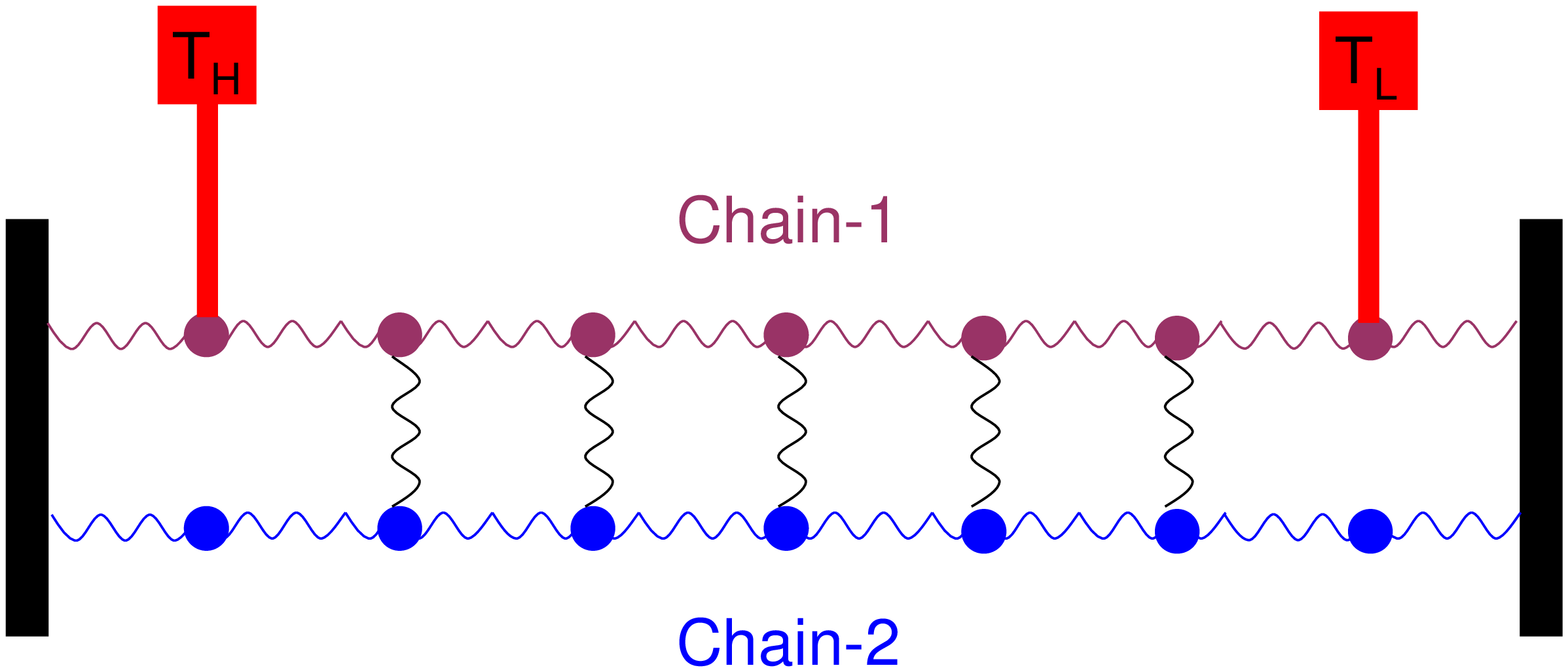}
\caption{ Schematic configuration of our approach to manipulate
thermal conductivity. Chain-1 and chain-2 represent the conductive
material and the substrate, respectively. The two particles on the
left and right ends of chain-1 are put into contact with thermostats
of temperature $T_{H}$ and $T_{L}$, respectively.}
\end{figure}

Fig. 1 shows the schematic configuration of our approach, which is
illustrated by a coupled atom chain model. The upside chain
(chain-1) represents the conductive material and the underside chain
(chain-2) represents the substrate. The two ends of chain-1 are
contacted with the thermostats with temperature $T_{H}$ and $T_{L}$,
respectively, while, chain-2 is free of thermostat contact. Each
node of chain-1, except for the two ends contacted with thermostat,
is coupled to the corresponding underside node of chain-2. This
ladder-like construction corresponds well to the real systems, where
the conductive materials are always fixed on some substrates.

Two coupled FPU chains are firstly considered to represent the approach in
details, with all the Hamiltonian parameters being in the reduced units.\cite%
{19,20} In the model, the atomic mass and spring constant of chain-1 ($%
m_{1},k_{1}$) are fixed to be 1.0, the anharmonic coefficient of chain-1 and
chain-2 ($\beta _{1}$, $\beta _{2}$) and the coupling strength $k_{c}$ are
all set to be 0.5. The atomic mass and spring constant of chain-2 $%
(m_{2},k_{2})$ are variables to manipulate thermal conductivity of
chain-1 (see Supporting Information, SI.I).

The nonequilibrium molecular dynamics (NEMD)\cite{19,20,22} method is used
to calculate the heat flux of chain-1 and chain-2 in the coupling system. It
is found that the phonon resonance between chain-1 and chain-2 plays an
important role on the thermal conductivity. Here we use the resonance angle $%
\Psi $=$\left\vert \arctan (\frac{\lambda _{2}}{\lambda
_{1}})\right\vert $ to describe the phonon resonance strength, where
$\lambda _{1}$ and $\lambda _{2}$ are the phonon amplitudes of
chain-1 and chain-2 after coupling (see Supporting Information,
 SI.II). It is obvious that the resonance
strength becomes a maximum (minimum) at $\Psi $ =$\frac{\pi }{4}$($\Psi $ = $%
0$).

\begin{figure}[tbp]
\includegraphics[scale=0.3,angle=-90]{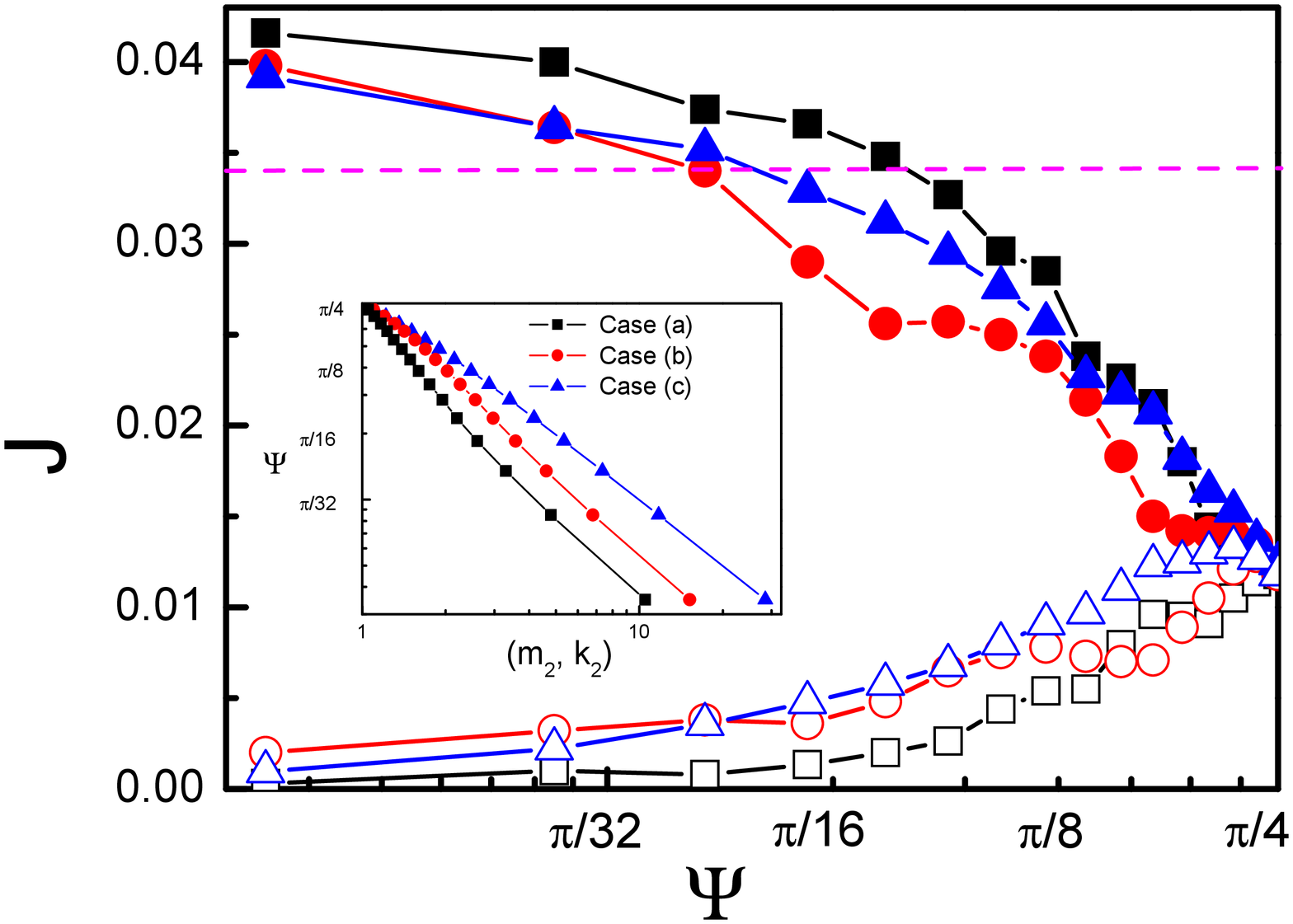}
\caption{ Heat flux of chain-1 and chain-2 with variation of
resonance angle $\Psi$ in the coupling system for three cases: (a)
$k_{2}$ =1, $m_{2}$ is varied; (b) $m_{2}$=1, $k_{2}$ is varied; (c)
both $m_{2}$ and $k_{2}$ are varied but keeping $m_{2}$=$k_{2}$. The
chain length $N$ is 50, and $T_{H}$=0.3, $T_{L}$=0.2. The solid
squares, triangles, and circles represent the heat flux of chain-1
for cases (a), (b), and (c), respectively; the corresponding open
ones represent heat flux of chain-2 for the three cases. The short
dashed line shows the heat flux of isolated chain-1. Chain-1's heat
flux monotonously decreases with $\Psi$ increasing, which can be
either larger or smaller than that of the isolated one depending on
the specific $\Psi $ value. The inset shows ($m_{2}$, $k_{2}$)
dependence of the resonance angle for the three cases. The resonance
angle is very sensitive to the ($m_{2}$, $k_{2}$) parameters.}
\end{figure}

Fig. 2 shows heat flux $J$ of chain-1 and chain-2 with variation of
the resonance angle. Since $\Psi $ is not uniquely determined by $m_{2}$ or $%
k_{2}$, many ($m_{2}$, $k_{2}$) sets can be corresponding to a $\Psi $
value, we mainly consider three cases: (a) $k_{2}$ =1, $m_{2}$ is varied;
(b) $m_{2}$=1, $k_{2}$ is varied; (c) both $m_{2}$ and $k_{2}$ are varied
but keeping $m_{2}$=$k_{2}$. For all the three cases, the heat flux of
chain-1 monotonously decreases with $\Psi $ increasing, while the heat flux
of chain-2 has a contrary behavior. Particularly, when the phonon resonance
becomes strong enough (with $\Psi $ nearby $\frac{\pi }{4}$), heat flux of
chain-1 gets much smaller than that of the isolated one ($J=0.034$),
suggesting a substantial reduction of thermal conductivity\cite{23}; when
the phonon resonance becomes small enough, however, the heat flux of chain-1
gets obviously larger than that of the isolated one, showing an increment of
thermal conductivity, which has not been explored so far. This interesting
phenomenon shows that the thermal conductivity of conductive material can be
efficiently manipulated through the substrate.

We note that the resonance angle is very sensitive to the ($m_{2}$,
$k_{2}$) parameters (Inset of Fig. 2). This suggests that one can
easily find a proper substrate to manipulate the thermal
conductivity in the application. For example, to decrease the
thermal conductivity, the substrate needs to have similar atomic
mass and spring constant with the conductive material; to increase
the thermal conductivity, the atomic mass and/or spring constant of
substrate only needs to be several times larger than that of
conductive material. Moreover, the heat-flux curves of cases (a),
(b), and (c) are very close to each other, indicating that the
thermal conductivity of chain-1 is insensitive to the specific type
of\ ($m_{2}$, $k_{2}$) compositions. Thus there would be rich
choices of substrate candidates for the thermal conductivity
manipulation. Consequently, the present approach can be easily
realized in the nanotechnologies, which provides a clear direction
for designing advanced nanoelectronic and thermoelectric devices.

The reduction of thermal conductivity after coupling can be
understood from the phonon-resonance effect, which induces strong
phonon scattering and thus reduces the thermal
conductivity.\cite{17,24,25,26,27} To understand the increment of
thermal conductivity, however, we need to invoke phonon band theory.
Here we consider an extreme condition, where the resonance angle is
zero and the increment of thermal conductivity is maximum. From the
coupling-harmonic-oscillator (CHO) model, the phonon dispersion of
chain-1
in the coupling system can be written as $\omega (q)=\sqrt{\frac{k_{1}}{m_{1}%
}}\cdot \sqrt{4\sin ^{2}(\frac{1}{2}q)+k_{c}}$. Compared with that of the
isolated case $\omega (q)=2\sqrt{\frac{k_{1}}{m_{1}}}\cdot \sin (\frac{1}{2}%
q)$, the phonon band of chain-1 has an obvious upshift after
coupling, the magnitude of which is proportional to $\sqrt{k_{c}}$
(Inset of Fig. 3).

\begin{figure}[tbp]
\includegraphics[scale=0.3,angle=-90]{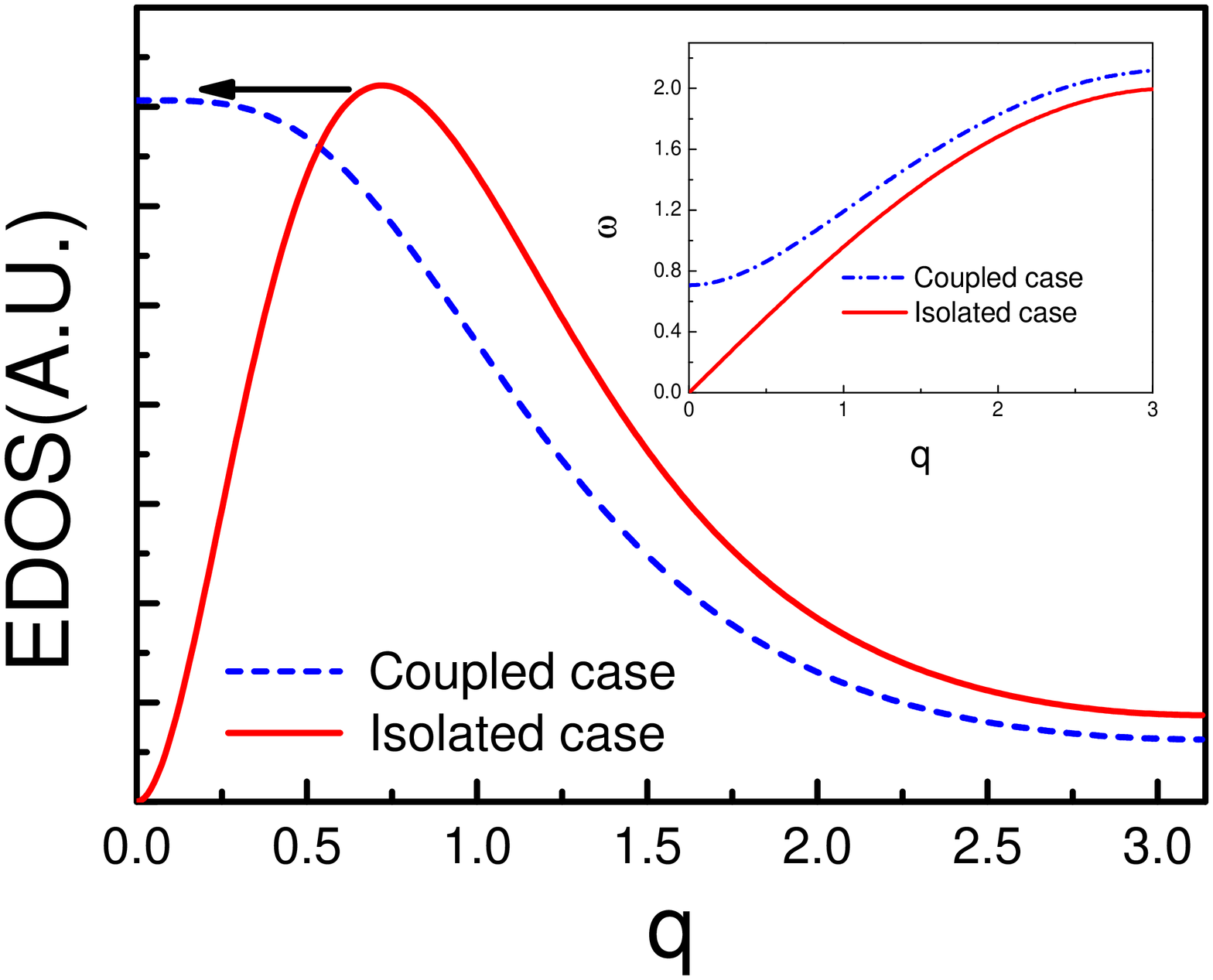}
\caption{PED distributions with the wave vector $q$ of chain-1  in
both the isolated and the coupled (with $\Psi =0$) cases at T=0.25.
The PED peak has an obvious shift to the small-$q$ direction after
coupling. The inset shows the corresponding phonon dispersions. }
\end{figure}

Based on the phonon dispersion, the phonon energy density (PED)\
distributions of chain-1 with wave vector $q$ can be further
calculated (see Supporting Information, SI.III). Compared with that
of isolated case, the PED peak has an obvious shift to the small-$q$
direction owning to the upshift of phonon band after coupling (Fig.
3), suggesting more energy has been carried and transported by the
small-$q$ phonons. The phonons' scattering power is proportional to
the phase difference between different atoms, which is hence
proportional to the $q$ value.\cite{28} More small-$q$ phonons being
responsible for the heat transport corresponds to smaller scattering
power, thus larger phonon mean free paths (PMFPs)\ in chain-1, which
increases the thermal conductivity (positive effect). On the other
hand, the upshift of the phonon band also results in less phonons
being excited for the heat transport, which in turn reduces the
thermal conductivity (negative effect). The thermal conductivity
variation induced by the phonon band upshift is attributed to such
two effects competing with each other (phonon-band-upshift effect).

\begin{figure}[tbp]
\includegraphics[scale=0.3,angle=-90]{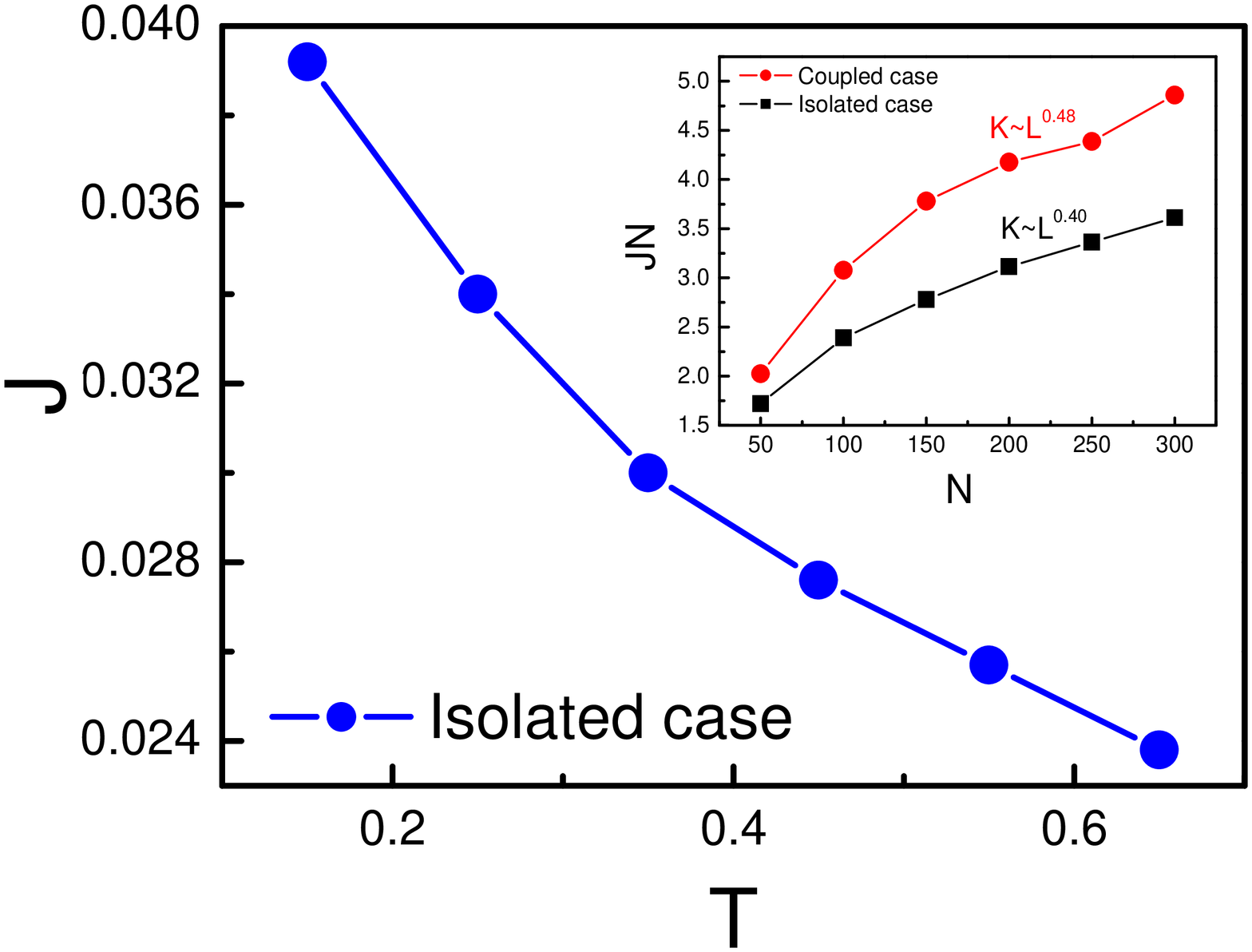}
\caption{(a) Temperature dependence of heat flux of isolated chain-1 ($N$%
=50). The phonon scattering effect has become dominant at our
simulation temperature (T=0.25). The inset shows chain length $N$
dependence
of total heat flux $JN$ of chain-1 in both the isolated and the coupled (with $\Psi=0$%
) cases. The larger fitting exponent implies that the increment of
thermal conductivity by the coupling gets more and more
distinguished with the chain length increasing.}
\end{figure}

We have also calculated the temperature dependence of heat flux of
isolated chain-1 (Fig. 4) and found the heat flux monotonically
decreases with the temperature increasing from T=0.15 to 0.65. It is
known that increasing temperature has two competitive effects on the
thermal conductivity: (i) It excites more high-frequency phonons
that enhance the thermal conductivity. (ii) It increases
phonon-phonon scattering that reduces the thermal
conductivity.\cite{13,20} Fig. 4 shows that effect (ii) has become
dominant at our simulation temperature (T=0.25), indicating that the
upshift of phonon band, which has similar effect with that of
decreasing temperature, has more positive effect than the negative
effect on the thermal conductivity. Hence the thermal conductivity
can be increased due to the phonon-band-upshift effect. Moreover,
inset of  Fig. 4 shows the chain length
dependence of total heat flux($JN$) of chain-1 in both the coupled (with $%
\Psi =0$) and isolated cases. As is shown, $JN$ of the isolated
chain-1 diverges as $JN\propto N^{0.4}$, consistent with the results
in previous reports.\cite{19,20} While, for the coupled case,
$JN\propto $ $N^{0.48}$. The larger fitting exponent implies that
the increment of thermal conductivity by the coupling gets more and
more distinguished with the chain length increasing.

Consequently, in the coupling system, the thermal conductivity
variation is owed to both the phonon-band-upshift effect and the
phonon-resonance effect that compete with each other. From the
results above (Fig. 2), the phonon-resonance effect can be easily
manipulated by changing the atomic mass and/or spring constant of
the substrate. Thus we can efficiently manipulate the thermal
conductivity of conductive material through coupling it to different
substrates.

\begin{figure}[tbp]
\includegraphics[scale=0.3,angle=-90]{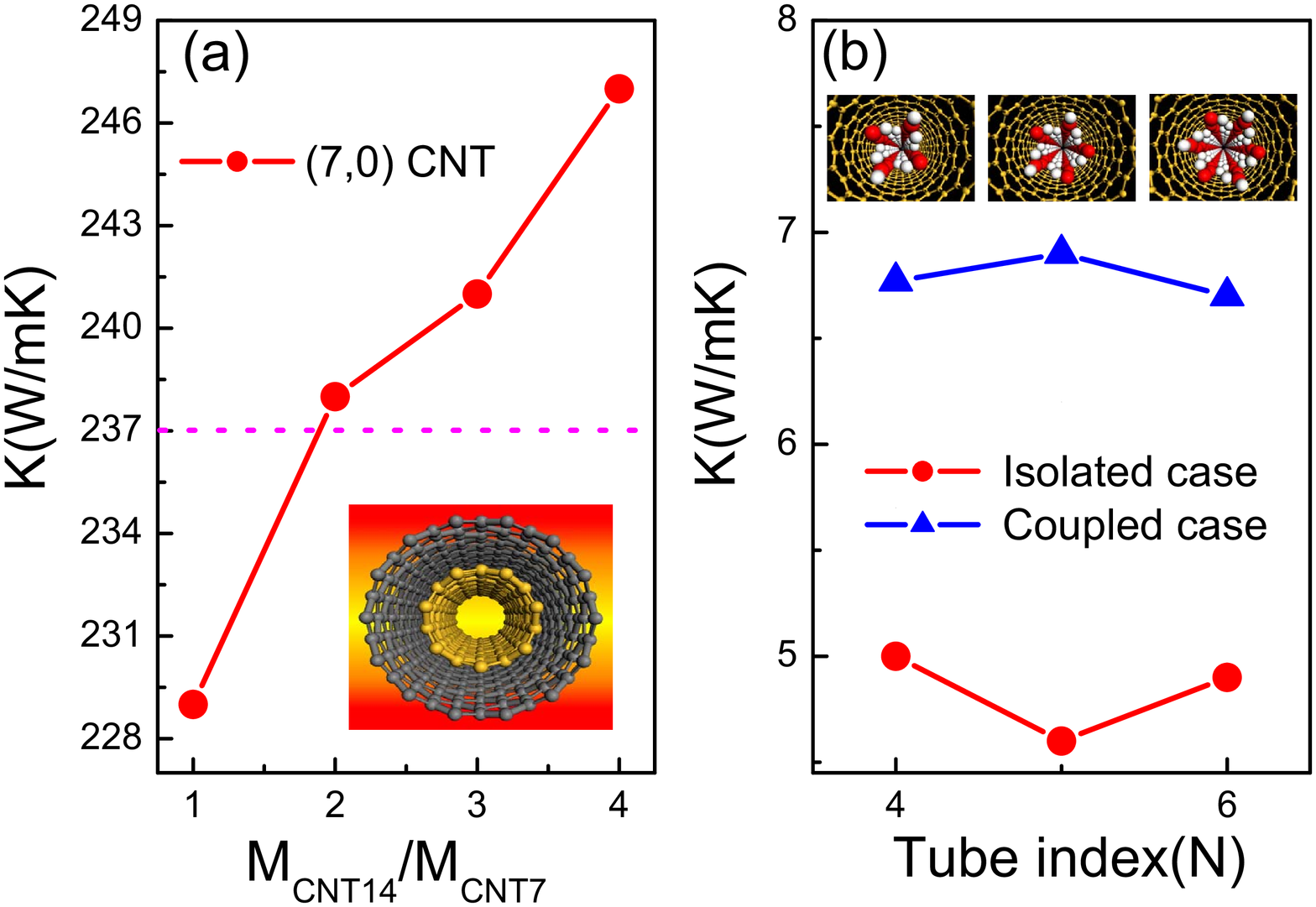}
\caption{(a) Thermal conductivity of (7,0) CNT coupled within (14,0)
CNT, with variation of (14,0) CNT's atomic mass $M_{CNT14}$
($M_{CNT7}$ is kept to be carbon-atom mass). The short dashed line
shows the thermal conductivity value of isolated (7,0) CNT. The
thermal conductivity of (7,0) CNT can be either substantially
decreased or increased depending on the value of $M_{CNT14}$. The
inset shows the structure of (7,0)@(14,0) DWNT. (b) Thermal
conductivity of Ice-NTs. Thermal conductivity of Ice-NTs has an
obvious increment after coupling. The inset shows optimized
structures of ice-NTs coupled within CNTs.}
\end{figure}
As a demonstration of this approach in the real systems, we have
calculated the thermal conductivity of a (7,0) carbon nanotube (CNT)
coupled within (14,0) CNT as a substrate((7,0)@(14,0) DWNT, see
Supporting Information, SI.IV). We set the atomic mass of (14,0) CNT
($M_{CNT14}$) as parameter to see how the thermal conductivity of
(7,0) CNT changes with the resonance angle. In the coupling system,
the resonance angle $\Psi $ and thus the thermal conductivity can be
efficiently manipulated through changing $M_{CNT14}$. As shown in
Fig. 5(a), thermal conductivity of (7,0) CNT obviously increases
with $M_{CNT14}$ increasing from one to
several times that of atomic mass of (7,0) CNT (carbon-atom mass, $M_{CNT7}$%
) owing to the reduction of phonon-resonance effect. Compared with
that of the isolated case, the thermal conductivity of (7,0) CNT can
be either substantially decreased or increased depending on the
value of $M_{CNT14}$, consistent with the results of coupled FPU
chain model, which can be well
understood form the coupling mechanism discussed above. When $%
M_{CNT14}=M_{CNT7}$ ($\Psi \approx \frac{\pi }{4}$), the
phonon-resonance effect that decreases the thermal conductivity is
dominant and the thermal conductivity is decreased by coupling. With
$M_{CNT14}$ increasing, the
phonon-resonance effect becomes more minor. When $M_{CNT14}\geq 2M_{CNT7%
\text{ }}(\Psi <\frac{\pi }{32}$)\cite{29}, the phonon-band-upshift
effect that increases the thermal conductivity becomes dominant, and
thus the thermal conductivity is increased by coupling.

Ice-NT coupled within a CNT can be considered as another realistic
illustration for our approach, where the Ice-NT and CNT correspond
to the conductive material and the substrate. Since the CNTs have
much larger spring constant than the Ice-NTs (from Supporting
Information,$\Psi \sim 0 $), an increment of thermal conductivity of
Ice-NTs is expected after coupling with CNTs. Fig. 5(b) shows the
calculated thermal conductivity of various Ice-NTs both with and
without CNT coupling (see Supporting Information, SI.V). As one can
see, the thermal conductivity of Ice-NTs has an obvious increment
after coupling, which is independent on the specific tube types.
These realistic illustrations further confirm the feasibility of our
approach for applications.

In summary, we have proposed a new approach to manipulate the
thermal conductivity. By coupling with different substrates, thermal
conductivity of the conductive nanomaterial can be either remarkably
decreased or increased, which can be realized in the device
applications. To decrease the thermal conductivity, the substrate
needs to have similar atomic mass and spring constant with the
conductive material; to increase the thermal conductivity, the
atomic mass and/or spring constant of the substrate needs to be
several times larger than that of the conductive material. Through
the illustrations of double-walled carbon nanotubes and coupled ice
nanotubes, we have further shown that this approach is applicable in
the real systems. Compared with the conventional treatments which
only reduce the thermal conductivity, our approach can truly realize
the thermal conductivity manipulation in solid nanomaterials without
destroying their structures.

 ~~~~~~\newline {\Large \textbf{Acknowledgments}:}

We gratefully thank Z. Y. Zhang and S. H. Wei for fruitful
discussions. This work was partially supported by the Special Funds
for Major State Basic Research, National Science Foundation of
China, Ministry of education and Shanghai municipality. The
computation was performed in the Supercomputer Center of Shanghai,
the Supercomputer Center of Fudan University.

\section{Supporting Information}

 "Manipulating thermal conductivity through substrate coupling",
 Zhixin Guo, Dier Zhang, and Xin-Gao Gong

\textbf{SI.I } Hamiltonian of the coupled FPU chains

The Hamiltonian of the coupled FPU chain model can be written as:

\begin{equation}
H=H_{1}+H_{2}+H_{c},
\end{equation}

\begin{equation}
H_{n}=\underset{i}{\sum }\left[ \frac{(p_{n}^{i})^{2}}{m_{n}}+\frac{1}{2}%
k_{n}(x_{n}^{i}-x_{n}^{i-1})^{2}+\frac{1}{4}\beta
_{n}(x_{n}^{i}-x_{n}^{i-1})^{4}\right] ,(n=1,2),
\end{equation}

\begin{equation}
H_{c}=\underset{i}{\sum }\left[ \frac{1}{2}k_{c}(x_{2}^{i}-x_{1}^{i})^{2}%
\right] ,
\end{equation}%
where $H_{1},H_{2},$ and $H_{c}$ are the Hamiltonian of chain-1,
chain-2, and the coupling term, respectively. $x_{n}^{i}$ ,
$p_{n}^{i}$ represent the displacement from the equilibrium
position, the momentum of the $ith$ particle; and $m_{n}$, $k_{n}$,
$\beta _{n}$ are the mass, spring constant and the anharmonic
coefficient of chain-$n$ $(n=1,2)$.

\textbf{SI.II} Definition of resonance angle $\Psi $

Using the coupling-harmonic-oscillator  (CHO) model, we can
calculate the phonon resonance strength. The amplitude ratio of
chain-2 and chain-1 after coupling can be expressed as:

\begin{equation}
\frac{\lambda _{2}}{\lambda _{1}}=\sqrt{\frac{m_{1}}{m_{2}}}\tan
(\alpha )\ (m_{2}\geq m_{1}),
\end{equation}

where $\lambda _{1}$ and $\lambda _{2}$ are the amplitudes of
chain-1 and chain-2, respectively. Here $\tan (\alpha
)=\frac{(\omega _{1}-\omega
_{2})\pm \sqrt{(\omega _{1}-\omega _{2})^{2}+4\omega _{c}^{2}}}{2\omega _{c}}%
\in \lbrack -1,1]$, and $\omega _{c}=$ $\frac{k_{c}}{\sqrt{m_{1}m_{2}}},$ $%
\omega _{1}=$ $\frac{k_{1}+k_{c}}{m_{1}},\omega _{2}=$ $\frac{k_{2}+k_{c}}{%
m_{2}}$, respectively. Then the resonance angle is defined as $\Psi
=\left\vert \arctan (\frac{\lambda _{2}}{\lambda _{1}})\right\vert $. When $%
\Psi =0,$ the phonon dispersion of chain-1 in the coupling system
can be written as:

\begin{equation}
\omega (q)=\sqrt{\frac{k_{1}}{m_{1}}}\cdot \sqrt{4\sin ^{2}(\frac{1}{2}%
q)+k_{c}},
\end{equation}
Thus the phonon band of chain-1 has an obvious upshift after
coupling, the magnitude of which is proportional to $\sqrt{k_{c}}$.

\textbf{SI.III} Calculation of PED distribution with wave vector

\begin{figure}[htbp]
\includegraphics[scale=0.3,angle=-90]{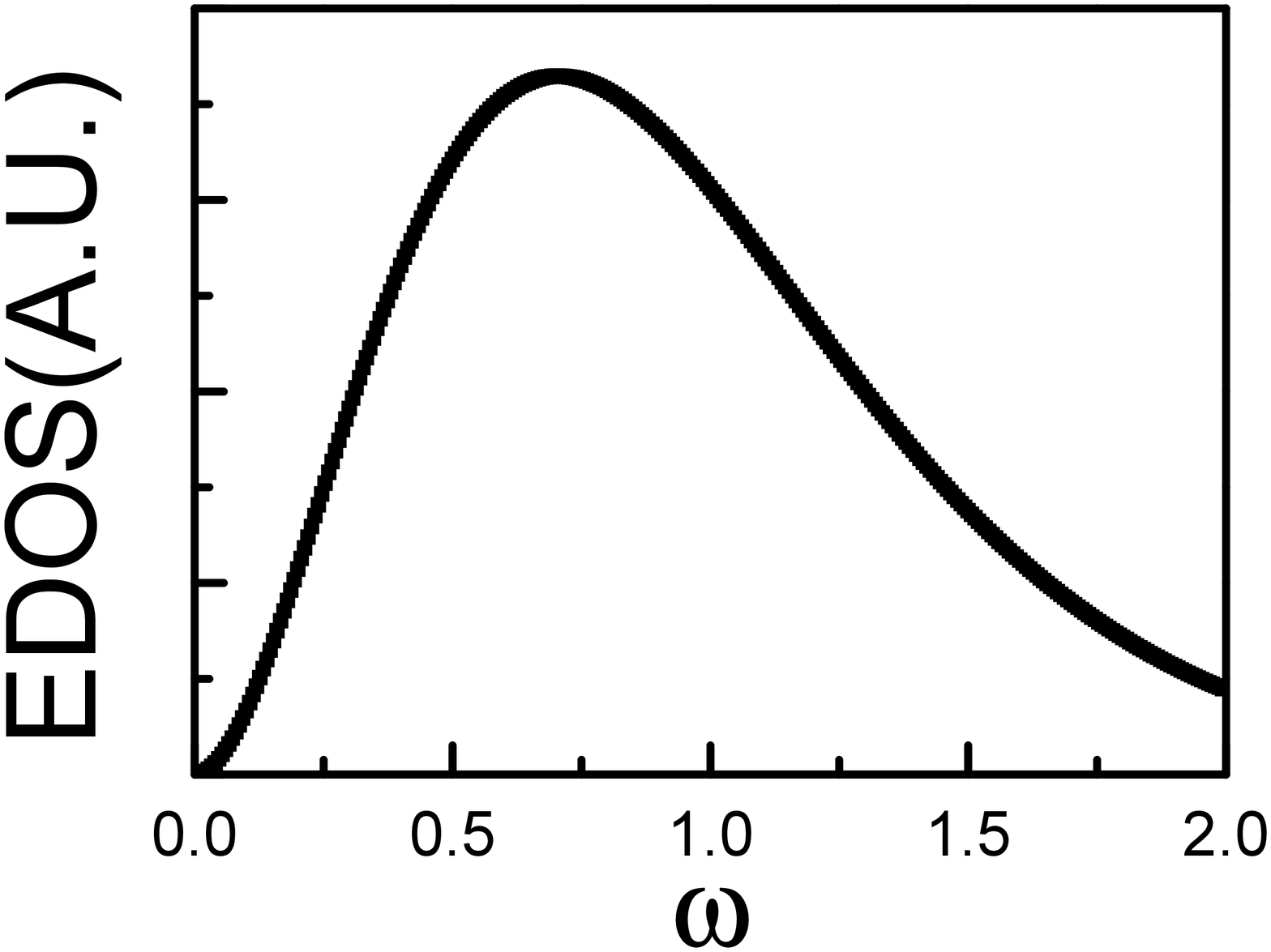}
\caption{PED distributions with phonon frequency $\omega $ of
isolated chain-1 at T=0.25.}
\end{figure}

The $q$ dependence of PED has an expression of $u(q)=\hbar \omega
(q)\times \eta \lbrack \omega (q),T], $with $\eta \lbrack \omega
(q),T]$ representing
the phonon distribution function at temperature $T$ for a certain frequency $%
\omega $. Since the phonon distribution in a heat-transport system
is mainly determined by the heat source that supplies phonons. We
suppose the conductive chain is coupled to an ideal heat source that
supplies phonons like the black-body, which can be represented by
the Nos\'{e}-Hoover thermostat\cite{1,2}. Then $\eta (\omega ,T)$
can be written as:

\begin{equation}
\eta (\omega ,T)=\frac{8\pi \omega ^{2}}{c^{3}}\frac{1}{e^{\hbar
\omega /k_{b}T}-1},
\end{equation}

\ Using equation (6), we also calculated the PED distributions with
phonon frequency $\omega $ of the isolated chain-1. As is shown in
Fig. 1, the result is consistent with that calculated by Li et al.
using the
numerical method, which confirms the reliability of our consideration.\cite%
{3}

\textbf{SI.IV} NEMD simulation for the double-walled carbon nanotube

In the simulation, the Tersoff potential was used to the describe
the carbon-carbon interaction\cite{4}, and the coupling interaction
between two CNTs was described by the LJ potential,\cite{5} which
was truncated at 10 \AA\ by a switching function. The simulated
(7,0)@(14,0) DWNT has a length of 121 \AA , containing 2352 atoms.
The wall-thickness of CNT was chosen to be 1.44 \AA\ for the
calculation of cross-section area.\cite{6,7} Fixed boundary
condition was applied along the axial direction of CNTs, where the
outmost two layers of each head were fixed.\cite{7,8} Then two
layers of each end of (7,0) CNT were put into contact with the
Nos\'{e}-Hoover thermostat with temperatures 310 and 290 K,
respectively, while the (14,0) CNT was free of thermostat contact.
To integrate equations of motion, the velocity Verlet method was
employed with a fixed time step of 1 $fs$. All results were obtained
by averaging about $10\ ns$ after a sufficient long transient time
($10\ ns$) to set up a nonequilibrium stationary state.

\textbf{SI.V} NEMD simulation for the Ice-NTs

Water-water intermolecular interaction was described by the
TIP4P\cite{9} potential and carbon-carbon interaction was described
by the Tersoff\cite{4} potential. As for the coupling term, the
CNT-water interaction was described by a carbon-oxygen LJ
potential.\cite{10} All the pair interactions were truncated at 10
\AA\ by a switching function. The simulation box
has a length of 121 \AA\ and the total number of water molecules inside is 44%
$\times $n for the (n,0) Ice-NT, where n = 4, 5, and 6. The Ice-NTs'
wall-thickness was chosen to be 2.75 \AA\ for the calculation of
cross-section area\cite{11}. Fixed boundary condition was applied
along the axial direction of Ice-NT/CNT, where the outmost two
layers of each head were fixed.\cite{7,8} Then two layers of each
end of Ice-NT were put into contact with the Nos\'{e}-Hoover
thermostat with temperatures 110 and 90 K, respectively, while the
CNT was free of thermostat contact. To integrate equations of
motion, the velocity Verlet method was employed with a fixed time
step of 1 $fs$. All results were obtained by averaging about $5\ ns$
after a sufficient long transient time ($5\ ns$) to set up a
nonequilibrium stationary state.


\begin{references}
\bibitem{23} Since the temperature difference is fixed to be 0.1, from the Fourier law, the thermal
conductivity is proportional to the heat flux value.
\end{references}
\end{document}